# Three-dimensional ultrasound-based online system for automated ovarian follicle measurement

**Authors:** Pedro Royo, M.D., Ph.D.[a,i], Elkin Muñoz, M.D.[b,i], José-Enrique Romero, Ph.D.[a,c], José-Vicente Manjón, Ph.D.[c], Catalina Roig, M.D.[d,i], Carmen Fernández-Delgado, M.D.[e,i], Nuria Muñiz, M.D.[f,i], Antonio Requena, M.D., Ph.D.[g,i], Nicolás Garrido, Ph.D.[i], Juan Antonio García-Velasco, M.D., Ph.D.[g,i], Antonio Pellicer, M.D., Ph.D.[h,i]

a. IVIRMA Global Research Alliance, Fertoolity, Valencia, Spain
b. IVIRMA Global Research Alliance, IVI RMA Vigo, Reproductive Medicine Unit, Spain
c. Instituto de Aplicaciones de las Tecnologías de la Información y de las Comunicaciones Avanzadas (ITACA), Universidad Politécnica de Valencia, Spain
d. IVIRMA Global Research Alliance, IVI RMA Palma de Mallorca, Reproductive Medicine Unit, Spain
e. IVIRMA Global Research Alliance, IVI RMA Bilbao, Reproductive Medicine Unit, Spain
f. IVIRMA Global Research Alliance, IVI RMA Lérida, Reproductive Medicine Unit, Spain
g. IVIRMA Global Research Alliance, IVI RMA Madrid, Reproductive Medicine Unit, Spain
h. IVIRMA Global Research Alliance, IVI RMA Rome, Reproductive Medicine Unit, Italy
i. IVIRMA Global Research Alliance, IVI Foundation, Instituto de Investigación Sanitaria La Fe (IIS La Fe)

**Short running title:** OSIS Ovary

**Corresponding Author:**

Pedro Royo

M.D., Ph.D.

Calle Colón, 1, 4ª planta (46004) Valencia (Spain)

+34 963 173 610

pedro.royo@ivirma.com

**Abstract**

**Background:** Ultrasound follicle tracking is an important part of cycle monitoring. OSIS Ovary (Online System for Image Segmentation for the Ovary) has been conceived aiming to aid the management of the workflow in follicle tracking, one of the most iterative procedures in cycle monitoring during ovarian stimulation.

In the present study, we compared OSIS Ovary (as three-dimensional ultrasound-based automated system) with the two-dimensional manual standard measurement method, in order to assess the reliability of the main measurements obtained to track follicle growth during ovarian stimulation cycles, the follicle size and count.

**Results:** Based on the mean follicle diameter and follicle count values obtained, the Pearson/intraclass correlation coefficients were 0.976/0.987 and 0.804/0.889 in ≥10mm follicles, 0.989/0.994 and 0.809/0.867 in ≥13mm follicles and 0.995/0.997 and 0.791/0.840 in ≥16mm follicles. The mean difference (MnD) for the mean diameter and follicle count was, respectively, 0.759/0.161 in ≥10mm follicles, 0.486/1.033 in ≥13mm follicles and 0.784/0.486 in ≥16mm follicles. The upper and lower limits of agreement (ULA and LLA) were 3.641/2.123 and 5.392/3.070 in ≥10mm follicles, 3.496/2.522 and 4.285/2.218 in ≥13mm follicles, and 3.723/2.153 and 2.432/1.459 in ≥16mm follicles. The limits of agreement range (LoAR) were 5.764/8.462 in ≥10mm follicles, 6.048/6.503 in ≥13mm follicles and 5.876/3.891 in ≥16mm follicles. $P<0.05$ was considered for all calculations

**Conclusion:**

As three-dimensional ultrasound-based automated system in comparison with two-dimensional manual method standard, we found OSIS Ovary as a reliable tool to track follicle growth during ovarian stimulation cycles

**Introduction**

Ultrasound follicle tracking is an important part of cycle monitoring. When the number of follicles increases, the manual measurement process can become tedious, imprecise and can potentially compromise the clinical outcomes (1,3,4,5). The use of automated volumetric software for follicle monitoring makes it possible to overcome these manual limitations, regardless of the moment in the cycle when the metrics are obtained and assessed, providing flexibility for both clinicians and patients (1,3). At the same time, more objective metrics are obtained with automated volume segmentation than with manual follicle measurement (2).

During follicular size estimation, 3D ultrasound initially counts the number of anechoic voxels inside each segmented follicle before displaying the numeric values, regardless of the shape of the follicle, thus minimizing the risk of repeatedly measuring the same follicle or missing growing follicles (1,2). Ovarian follicles are considered spheroids (non-perfect spheres) due to interfollicular coaptation between the ovarian cortex and the stromal region **(Figure 1)** and/or operator-induced compression of the ovarian surface with the transvaginal probe. To overcome this technical limitation (and its impact on follicular size estimations), the existing consensus for volumetric measurement supports using the calculation of the relaxed sphere diameter (dV) as a standard, instead of the estimation of the mean follicle diameter. This extrapolates 3D follicular volume to a perfect-shaped sphere, making it possible to calculate its diameter (dV) and avoid irregular follicle shaping as a confounding factor for measurement comparisons (1,3,4,5). 3D ultrasound tracking metrics, dV and follicle volume are equivalent ($v=4/3\pi r^3$) (1,2) and can be used for follicle monitoring, but this requires shifting from manual (2D) to automated (3D) methods (1,4,5).

OSIS Ovary (Online System for Image Segmentation for the Ovary) has been conceived aiming to aid the management of the workflow in follicle tracking, one of the most iterative procedures in cycle monitoring during ovarian stimulation (OS) (1,2,3).

The system drives patients' dataflow directly from any 3D ultrasound machine to OSIS Ovary cloud-based server, enabling an easier adoption in the clinical offices. The volumetric-based follicle tracking with OSIS Ovary, has been developed to contribute to the use of 3D ultrasound technology as a useful resource for reproductive monitoring (1,3,4).

The purpose of the present study was to compare automated OSIS Ovary system to standard conventional 2D manual ultrasound, before its use in clinical practice.

**Materials and Methods**

**-Reference population**

A prospective cohort study was performed among patients undergoing OS. They were invited to participate in the study between May 2020 and February 2021. All participants signed an informed consent form prior to the start of the study.

**-Inclusion criteria**

Consecutive patients aged 18 years or more undergoing OS for IVF, and/or oocyte vitrification.

**-Exclusion criteria**

Patients were excluded from the study if they cancelled their OS cycle or if the three required 2D and 3D ultrasound study datasets were not fully completed (measured, saved and processed).

**-OSIS Ovary Software Architecture**

The OSIS Ovary software system was designed to be accessed via web browser, and it can be used to perform automatic segmentation, 2D visualization, 3D rendering and ovary follicle quantification (number and size).

**-Library of manually delineated volumes**

We used 100 ovarian volume ultrasound test dataset, gathered in 2019 in IVI RMA Vigo from volunteer OS patients, to build a knowledge base that was used to train the OSIS Ovary segmentation algorithm. OSIS Ovary image engineers performed manual volume delineations to produce 100 ovarian follicle segmentation masks, as shown in **Supplemental Figure 1**. After this step, manual segmentations were examined using specialized software to reduce errors and increase consistency.

**-Segmentation model and training**

OSIS Ovary model consist of a deep convolutional neural network development (DCNN) based in the U-NET topology (6) modified to take volumetric input (see **Supplemental Figure 1**) and to perform automatic follicle segmentation. This network takes a 3D ultrasound volume rendering (volumetric network) for a single ovary as input and produces another 3D volume rendering with the same dimensions including a mask of the ovarian follicles. The DCNN model was trained using the 100 manually delineated cases trough several iterations aimed at maximizing the similarity between manual delineation and the automatic segmentation produced by the network.

The DCNN model was implemented in Python language (using Tensorflow and Keras libraries) consisting of a series of steps to reduce dimensionality over an anatomical 3D volume (contracting path) to produce a latent space with the relevant characteristics and a series of steps to increase back dimensionality from the latent space- (expansive path) to produce a segmentation of the 3D volume anatomy. Each step in the contracting path of the U-NET consists of a 3D convolution layer composed by 32 (64 at the second level, 128 at the third level and 256 at the laten space) filters followed by a batch normalization layer and a rectified linear unit (ReLU). This is repeated three times and finishes replacing the batch normalization layer by a max pooling layer (no drop-out is used to train the model). The expansive path consists of the same layer structure decreasing the number of filters by a factor 2 for upper levels and replaces the max pooling layer by an upsampling layer based in a trilinear filter with no trainable parameter.

In addition to the classic U-NET architecture we introduced residual connections that concatenate the output of each step of the contracting path to the corresponding level in the expansive path. Also, during the training, the output of each expansive path step was considered to compute the loss, which is a logarithmic mean DICE (7). We used the DICE coefficient to measure segmentation similarity and, to increase the accuracy of our method, added a post-processing step after neural network segmentation to detect possible merged follicles and separate them. To do so, we use geodesic distance maps to calculate each follicle center, called seed, and grow a region from each seed (**Supplemental Figure 2**).

The optimum mixing factor to weight the different level outputs in the loss calculation was 0.0 for the deepest level, 0.3 for the middle one and 0.7 for the shallowest level. The number of filters used in each layer was sent up to its possible maximum value considering the limitations of memory of the nVidia DGX1 server (NVIDIA Corporation) used to this purpose. The optimizer used to train the method was Adam algorithm, with the default learning rate of 0.001. To feed the generator for training we found positive to use data augmentation we called "MixUp" by generating on-the-fly a linear combination of two training inputs with weight of 0.3 and 0.7 for each image. The model was trained using a 10-fold strategy already explained performing 300 epoch for each fold. The optimal performance (mean value of 10 folds) was an overall dice of 0.92, a correlation for the measured volume of 0.99 and a correlation for the number of follicles detected of 0.95.

**-Follicle metrics delivery**

OSIS Ovary provides the follicular measurements (number of follicles and relaxed sphere diameter (dV, in mm) based on the automated volumetric segmentations. Before calculating the metrics, morphological operations are performed to improve the segmentation. These operations are aimed to separate possible touching follicles identified as a single one. These operations are not part of the AI model output and consist of calculate each follicle center called seed using geodesic distance maps and perform a region growing from each seed (**Supplemental Figure 2**).

The process to obtain the metrics is the following: the number of follicles consists of counting the number of isolated groups of voxels (classified as follicle with class 1) in the segmentation; subsequently, the volume was calculated as the sum of all the voxels from a single follicle and multiplying by voxel volume. Voxel volume is given by input image resolution, and the relaxed sphere diameter (dV) is given the measured volume of a follicle, its dV corresponds to the diameter of a perfect sphere with the same volume.

**-Study protocol**

Three transvaginal (2D and 3D) ultrasounds scans were mandatory performed on volunteer patients on the three clinically relevant OS monitoring days: $4^{th}$- $5^{th}$ day (first follicular assessment), $6^{th}$ -$7^{th}$ day (introduction of the GnRH antagonist) and $8^{th}$-$9^{th}$ day (oocyte pickup scheduling).

Based on previous studies (4,8), it was estimated to include a minimum of 100 ultrasound datasets in the study.

**-Data acquisition**

All volunteers were transvaginal-via ultrasound examined by the investigators, using a Voluson™ P6 ultrasound scanner, mounted with 3D/4D RIC5-9Mhz transvaginal probes (General Electric Healthcare). Each subject was asked to empty the bladder and was then scanned in a supine position. A routine 2D ultrasound assessment of the pelvis was first performed to measure the follicles using the standard manual mode (described in the next section). The ovaries were then visualized in the longitudinal plane and the volume mode was activated. The 3D box defining the area of interest was then moved and adjusted and the sweep angle set to 90° to ensure that all of the ovary shape was included in the volume. The patient was asked to remain as still as possible, and every effort was made by the operator to limit inadequate movements of the transducer. A 3D dataset was then acquired using the Voluson™ maximum quality-slow sweep mode. The resultant multiplanar display was examined to ensure that the ovary had been captured in its entirety. Standardization of the ultrasound settings was assured by using the same predefined probe program (Voluson™ gynecology preset with gyn settings) without adjustment once the program had been loaded. All 2D and 3D scan volume datasets of the ovaries were saved, dated, and encrypted in the OSIS Ovary system server (9). The dataset used to validate OSIS Ovary consisted of a collection of 534 transvaginal 3D ultrasound volume datasets (from 89 consecutive volunteers). Each volume was obtained from each ovary of the patients during the required three examinations during each performed OS. Patient data were pseudonymized.

**-Study variables and ranges**

The individual mean follicle diameter (mm) per ovary was recorded. In 2D manual measurements, the mean follicle diameter is calculated by taking two orthogonal diameters, largest+smallest/2. OSIS Ovary, on the other hand, calculates the follicular volume and relaxed sphere diameter (dV) (1,4,5).

The mean follicle diameter ranges considered were in line with clinical criteria: ≥10 mm ($4^{th}$- $5^{th}$ day), ≥13 mm ($6^{th}$ -$7^{th}$ day of OS), and ≥16mm ($8^{th}$-$9^{th}$ day of OS).

**-Statistics**

Once manual and OSIS Ovary measurements were recorded, the results obtained were compared to assess whether OSIS Ovary could provide clinical benefits. The measures used to compare both metrics were Pearson's r correlation coefficient (Matlab™, Release 2022a) and the intraclass correlation coefficient (ICC), which represents absolute agreement with one-way random-effects model in R, version 4.1.3; r ≥0.8 and ICC ≥0.8 were considered indicators of good correlation and agreement (1,4,5,9). Bland-Altman plots (Matlab™, Release 2022a) were used to calculate the mean difference (MnD), upper and lower limits of agreement (ULA and LLA), and the limits of agreement range (LoAR) between manual and OSIS Ovary measurement methods (1,4,5). P<0.05 was considered for all calculations. Manual and OSIS Ovary measurements were expressed as average results (mean +/- SD).

**Results**

Average mean diameter and follicle count values were reported in **Table 1**. Based on the mean follicle diameter and follicle count values obtained, the Pearson/intraclass correlation coefficients were 0.976/0.987 and 0.804/0.889 in ≥10mm follicles, 0.989/0.994 and 0.809/0.867 in ≥13mm follicles and 0.995/0.997 and 0.791/0.840 in ≥16mm follicles. **Table 2** shows the mean difference (MnD) for the mean diameter and follicle count: 0.759/0.161 in ≥10mm follicles, 0.486/1.033 in ≥13mm follicles and 0.784/0.486 in ≥16mm follicles. The upper and lower limits of agreement (ULA and LLA) were 3.641/2.123 and 5.392/3.070 in ≥10mm follicles, 3.496/2.522 and 4.285/2.218 in ≥13mm follicles, and 3.723/2.153 and 2.432/1.459 in ≥16mm follicles. The limits of agreement range (LoAR) was 5.764/8.462 in ≥10mm follicles, 6.048/6.503 in ≥13mm follicles and

5.876/3.891 in ≥16mm follicles. The metrics contained in Table 2 were graphically depicted in Bland-Altman plots **(Figure 2)**.

**Discussion**

Ultrasound follicle tracking is a routine process during IVF treatment. The purpose of the present study was to test a research volumetric ultrasound system by comparing the measurements obtained using OSIS Ovary to manual outcomes (in terms of the number of follicles and their mean diameter). The results reflected in the study showed enough reliability for clinical use.

OSIS Ovary showed correlation and agreement with manual methods ≥0.9 in the mean diameter and ≥0.8 per ovarian follicle count; this remained stable across the three follicle growth stages **(Table 2)**. Both methods measured almost the same mean diameter and number of follicles. While MnDs and mean LoARs remained steady, reporting values of ≤1 and ≤6 respectively, a minimal deviation was detected only for smaller follicles (8.4 LoAR in ≥10mm) **(Table 2)**.

Although the follicle growth study stages did not match exactly, they were within the same clinical management ranges, with the main two studies comparing the manual method to SonoAVC™—conducted by Ata (10-13mm/14-17mm/≥18mm) (4) and Raine-Fenning (≥10mm/≥14mm/≥18mm) (5)—enhancing the comparability of our results (9). Based on manual follicle number agreement, OSIS Ovary provided a shorter mean LoAR (5.8, versus 8.2 (4) and 6.6 (5)) and also a steadier LoAR (5.76/6.04/5.87) than SonoAVC™ (12.8/6.8/4.9) (4), (7.02/7.5/4.98) (5). The MnD was slightly higher using OSIS Ovary (0.8, versus 0.6 (4) and 0.32 (5), with no more than one follicle remaining in all stages (1.16/1.03/0.48), as others have previously shown (-1.3/0.1/0.5) (4), (-0.5/-0.5/0.15) (5). Regarding the correlation between OSIS Ovary and manual data (r

0.80/0.81/0.79 vs 0.96/0.87/0.78) (4) as well as agreement (ICC 0.89/0.87/0.84 vs 0.91/0.88/0.84) (5), our system obtained very similar outcomes comparing manual and automated (using SonoAVC™) follicle counts in the three clinical stages of the series compared. Our dataset, comparable to other published studies (1,4,5,9), was focused on follicle-growth measuring comparison, assessing our research system with a view to performing future clinical use (1,4,5). Similarly, these follicle-tracking referenced studies, were also perfomed during OS cycles in unrestricted OS clinical scenarios (4,5,8), like OSIS Ovary was.

The objective and standardized metric for volume-derived outcomes used to compare OSIS Ovary and SonoAVC™ to the manual method was dV (1,2,4,5,10,11,12). Based in this measurement, the performance of OSIS Ovary can be considered enough to be used reliably.

Further in-depth clinical studies are being conducted using OSIS Ovary in different clinical scenarios, in order to shift it from clinical research field to clinical daily practice.

**Conclusion**

OSIS Ovary constitutes a reliable follicle-measurement tool, because the performance observed can be considered enough in comparison with the manual standard method. Any case, it must be progressively clinical studied to be assessed and optimized in the next future with the purpose of making it easier to track follicle growth during ovarian stimulation cycles.

**Acknowledgements**

We would like to express our sincere gratitude to Victoria de Uriarte for supporting us in the writing assistance, as well as Belen Fos and Marcos Alepuz for supporting us in the management of the publication.

OSIS Ovary is intended for use only by qualified healthcare professionals (nurses, image technicians or physicians) and is indicated for the treatment of adult patients, mainly undergoing ovarian stimulation. It can be used for multiple patients at the same time, and only in clinical environments. The study has been approved by the Institutional Review Boards of Pontevedra-Vigo-Ourense (Xunta de Galicia), 1712-PAM-121-PR (17.07.2018), and Hospital Universitario y Politécnico La Fe (Departament de Salut, Valencia), 1907-VGO-073-PR (15.01.2020).

The OSIS Ovary project (registered code IDI-20181216) was funded by the Research and Development Project Line of the CDTI (Centre for the Development of Industrial Technology, Ministry of Science and Innovation - Government of Spain) and co-funded by the Multiregional OP for Spain ERDF 2014-20 (European Regional Development Funds within the Multiregional Operational Programme for Spain - European Union).

All volunteers as participants signed an informed consent form prior to the start of the study.

All the volunteers and authors gave their agreement to the submission of this manuscript.

No conflict of interests declared.

**References**

[1] Vandekerckhove F, Bracke V, De Sutter P. The Value of Automated Follicle Volume Measurements in IVF/ICSI. Front Surg. 2014 May 28;1:18.

[2] Raine-Fenning N, Jayaprakasan K, Clewes J, et al. SonoAVC: a novel method of automatic volumen calculation. Ultrasound Obstet Gynecol. 2008;31(6):691-696

[3] Deutch TD, Joergner I, Matson DO, et al. Automated assessment of ovarian follicles using a novel three-dimensional ultrasound software. Fertil Steril. 2009 Nov;92(5):1562-1568.

[4] Ata B, Seyhan A, Reinblatt SL, Shalom-Paz E, Krishnamurthy S, Tan SL. Comparison of automated and manual follicle monitoring in an unrestricted population of 100 women undergoing controlled ovarian stimulation for IVF. Hum Reprod. 2011 Jan;26(1):127-133.

[5] Raine-Fenning N, Jayaprakasan K, Deb S, et al. Automated follicle tracking improves measurement reliability in patients undergoing ovarian stimulation. Reprod Biomed Online. 2009 May;18(5):658-663.

[6] Ronneberger O, Fischer P, Brox T. "U-Net: Convolutional Networks for Biomedical Image Segmentation". MICCAI2015, pp. 234-241, 2015.

[7] Milletari, F., Navab, N., Ahmadi, S.A. "V-Net: Fully Convolutional Neural Networks for Volumetric Medical Image Segmentation". Fourth International Conference on 3D Vision (3DV). pp. 565–571, 2016.

[8] Lamazou F, Arbo E, Salama S, Grynberg M, Frydman R, Fanchin R. Reliability of automated volumetric measurement of multiple growing follicles in controlled ovarian hyperstimulation. Fertil Steril. 2010 Nov;94(6):2172-2176.

[9] Jayaprakasan K, Walker KF, Clewes JS, Johnson IR, Raine-Fenning NJ. The interobserver reliability of off-line antral follicle counts made from stored three-dimensional ultrasound data: a comparative study of different measurement techniques. Ultrasound Obstet Gynecol. 2007 Mar;29(3):335-341.

[10] Raine-Fenning N, Jayaprakasan K, Chamberlain S, Devlin L, Priddle H, Johnson I. Automated measurements of follicle diameter: a chance to standardize? Fertil Steril 2009a; 91:1469 – 1472.

[11] Raine-Fenning N, Jayaprakasan K, Deb S, et al. Automated follicle tracking improves measurement reliability in patients undergoing ovarian stimulation. Reprod Biomed Online 2009b; 18:658 – 663.

[12] Gerris J, De Sutter P. Self-operated endovaginal telemonitoring (SOET): a step towards more patient-centred ART? Hum Reprod 2010; 25:562– 568.

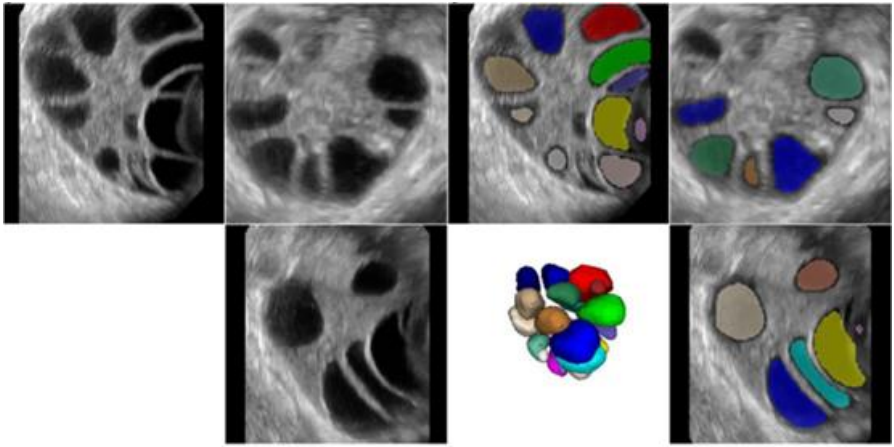

**Figure 1.** Three-dimensional ultrasound images of an ovarian volume and its segmentation.

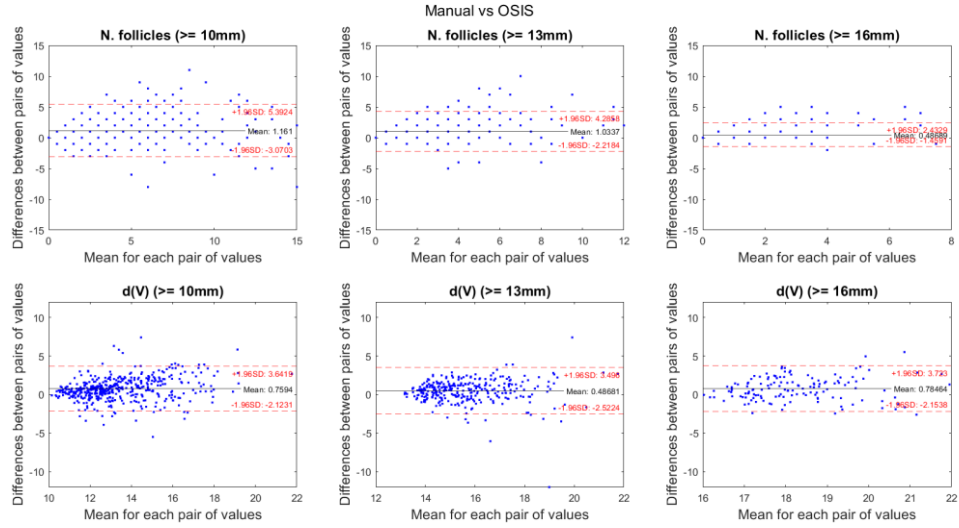

**Figure 2.** Bland-Altman plots showing the mean (x-axis) and difference (y-axis) between the two different methods used in the folliculometry: OSIS Ovary vs manual measurements. The mean difference (black full line) and 1.96 SD (red dashed lines) are shown.

|  | ≥10 mm | | ≥13 mm | | ≥16 mm | |
|---|---|---|---|---|---|---|
|  | OSIS | Manual | OSIS | Manual | OSIS | Manual |
| Mean diameter (mm) | 13.18±2.57 | 13.84±3.09 | 15.53±2.17 | 15.88±2.80 | 17.98±1.93 | 18.43±2.93 |
| Follicles per ovary | 2.32±3.21 | 2.73±3.72 | 2.47±1.90 | 3.60±2.72 | 1.73±0.96 | 2.36±1.50 |

**Table 1**. Average results for OSIS Ovary and manual measurements (mean +/- SD)

|  | ≥10 mm | | | ≥13 mm | | | ≥16 mm | | |
| --- | --- | --- | --- | --- | --- | --- | --- | --- | --- |
|  | r | ICC |  | r | ICC |  | r | ICC |  |
| **Mean diameter (mm)** | 0.976 | 0.987 |  | 0.989 | 0.994 |  | 0.995 | 0.997 |  |
| MnD |  |  | 0.759 |  |  | 0.486 |  |  | 0.784 |
| ULA+ |  |  | 3.641 |  |  | 3.496 |  |  | 3.723 |
| LLA- |  |  | 2.123 |  |  | 2.522 |  |  | 2.153 |
| LoaR |  |  | 5.764 |  |  | 6.048 |  |  | 5.876 |
| **Follicles per ovary** | 0.804 | 0.889 |  | 0.809 | 0.867 |  | 0.791 | 0.840 |  |
| MnD |  |  | 1.161 |  |  | 1.033 |  |  | 0.486 |
| ULA+ |  |  | 5.392 |  |  | 4.285 |  |  | 2.432 |
| LLA- |  |  | 3.070 |  |  | 2.218 |  |  | 1.459 |
| LoaR |  |  | 8.462 |  |  | 6.503 |  |  | 3.891 |

*p-values less than 0.05*

**Table 2.** Pearson Correlation Coefficient ® and Intraclass Correlation Coefficient (ICC), Mean Difference (MnD), Upper and Lower Limits of Agreement (ULA and LLA), Limits of Agreement Range (LoAR) between OSIS Ovary and manual measurements considering +/- 1.96 SD

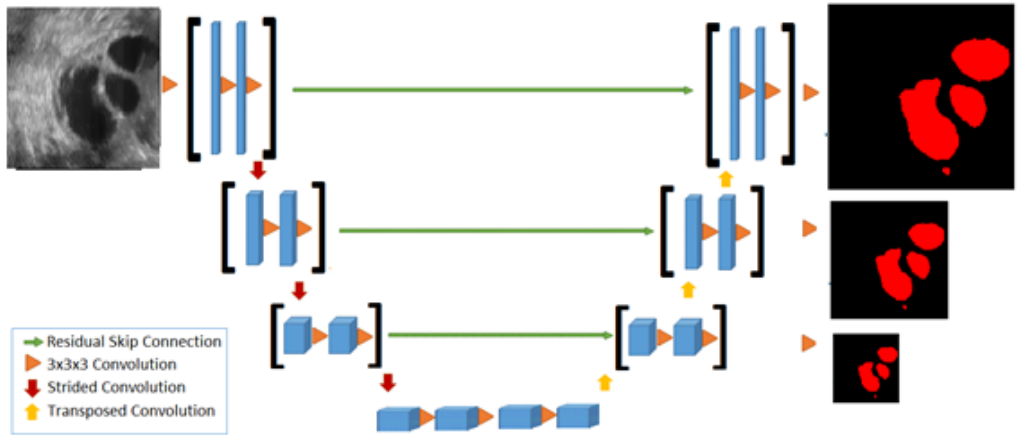

**Supplemental Figure 1:** UNET 3D topology

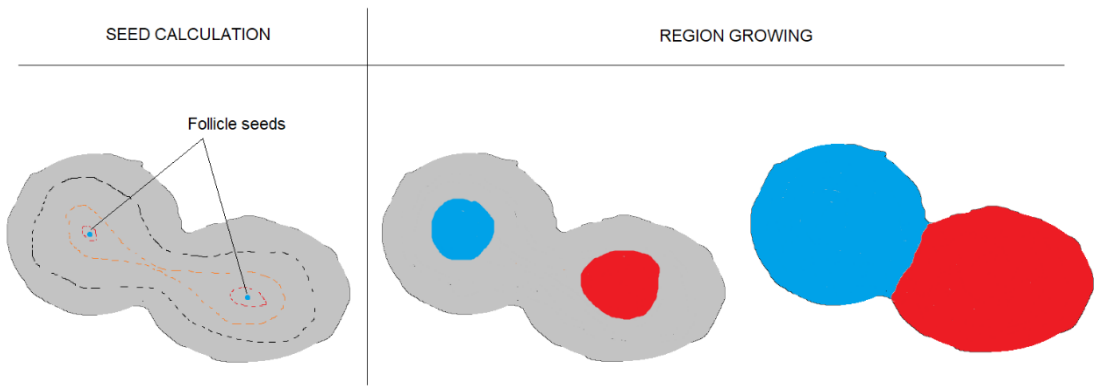

**Supplemental Figure 2:** Follicle separation